\documentclass[conference]{IEEEtran}
\usepackage{cite}
\usepackage{amsmath,amssymb,amsfonts}
\usepackage{algorithmic}
\usepackage{graphicx}
\usepackage{textcomp}
\usepackage{xcolor}
\def\BibTeX{{\rm B\kern-.05em{\sc i\kern-.025em b}\kern-.08em
    T\kern-.1667em\lower.7ex\hbox{E}\kern-.125emX}}

\begin{document}

\title{Preliminary Analysis of Skywave Effects on MF DGNSS R-Mode Signals During Daytime and Nighttime}

\author{\IEEEauthorblockN{Suhui Jeong} 
\IEEEauthorblockA{\textit{School of Integrated Technology} \\
\textit{Yonsei University}\\
Incheon, Republic of Korea \\
ssuhui@yonsei.ac.kr}
\and
\IEEEauthorblockN{Pyo-Woong Son${}^{*}$}
\IEEEauthorblockA{\textit{Korea Research Institute of Ships and Ocean Engineering} \\
Daejeon, Republic of Korea \\
pwson@kriso.re.kr} 
{\small${}^{*}$ Corresponding author}
}

\maketitle

\begin{abstract}
Accurate positioning, navigation, and timing (PNT) performance are prerequisites for several technologies today. 
In a marine environment, it is difficult to visually identify one’s position accurately, leading to safety concerns. 
Currently, PNT information is provided mainly from Global Navigation Satellite Systems (GNSS); however, it is vulnerable to radio frequency interference, spoofing, and ionospheric anomaly. 
Therefore, research on a backup system is needed. 
Ranging Mode (R-Mode), a terrestrial integrated navigation system, is being investigated for use in a marine environment.
R-Mode is a positioning technology that integrates terrestrial signals of opportunity such as medium frequency (MF) differential GNSS (DGNSS), very high frequency (VHF) automatic identification system (AIS), and enhanced long-range navigation (eLoran) signals. 
Previous studies in Europe show that signals in the MF band differ greatly in accuracy between daytime and nighttime. 
This difference is primarily caused by skywave. 
In this study, the MF DGNSS R-Mode signal transmitted from Chungju, Korea was received in Daesan and Daejeon, Korea. 
The skywave effect during daytime and nighttime was compared and investigated. 
In addition, the continuous wave intensity of the R-Mode signal was increased during the nighttime to compare its effect on the measurement accuracy.
\end{abstract}

\begin{IEEEkeywords}
MF DGNSS R-Mode, skywave, PNT
\end{IEEEkeywords}

\section{Introduction}

Accurate positioning, navigation, and timing (PNT) information is one of the essential requirements of current technologies, such as telecommunications, transportation, port logistics, and numerous location-based applications. 
Owing to the lack of visually identifiable features in the marine environment, it is difficult to identify location accurately; therefore, safety concerns dictate the importance of PNT information. 
Accurate PNT information is necessary for contemporary marine technologies such as autonomous ships. 
PNT information are currently obtained, both on land and at sea, mainly using Global Navigation Satellite Systems (GNSS) \cite{Enge1994:global, Park20:Effects, Kim18:Low, Lee22:Urban}. 
However, the GNSS system relies on satellite transmission; hence, the received signal is very weak. 
Therefore, it is vulnerable to artificial interference such as jamming and spoofing \cite{Park21:Single, Park18:Dual, Kim19:Mitigation, Schmidt20} or ionospheric anomaly \cite{Lee22:Optimal, Sun21:Markov}. 
Several backup navigation systems have been studied \cite{Li20, Jia21:Ground, Jeong20:RSS, Han19:Smartphone, Lee20:Integrity, Rhee19:Low, Rhee18:Ground, Kim17:SFOL, Lee22:SFOL, Shin17:Autonomous, Kim17:Simulation, Lee22:Evaluation, Kang21:Indoor}. 
A navigation system using ``Signals of Opportunity'' (SoOp), is a system that utilizes current broadcast radio frequency (RF) signals for navigation, either directly or after certain changes \cite{Mcellroy2006:navigation,Wang21, Huang22, Yang22}. 
Research on ``Ranging Mode'' (R-Mode) navigation systems using medium frequency (MF) signals for differential GNSS (DGNSS) transmission and very high frequency (VHF) signals for automatic identification system (AIS) transmission as SoOp is being conducted in the maritime sector \cite{Johnson2014:feasibility, Johnson2014:feasibility1, Johnson2014:feasibility3, Johnson2017:initial, Johnson2020:R-Mode, Swaszek2012:ranging, Jeong21:Development}. 
R-Mode investigation using three different configurations of MF DGNSS, VHF AIS, and enhanced long-range navigation (eLoran) \cite{Son19:Universal, Son18:Novel, Kim22:First, Rhee21:Enhanced, Williams13} are also being conducted \cite{Johnson2014:feasibility1}. 
Currently, the R-Mode technology development in Korea is underway \cite{Han2021:R-Mode}.

The maritime DGNSS ground reference station transmits GNSS correction information by minimum shift keying (MSK) modulation of a signal in the 285--325 kHz frequency band. 
In the R-Mode, continuous waves (CW) can be added to the existing MF DGNSS signals to provide positioning capability. 
Validation of the MF R-Mode in the North Sea area revealed a significant difference in accuracy, approximately 10 times, between daytime and nighttime positioning  \cite{Johnson2014:feasibility3}. 
The major cause of error was identified as skywave. 
According to \cite{Johnson2014:feasibility3}, skywaves occur after sunset and have a significant effect on the positioning accuracy at night. 
Unlike the pulse signal of eLoran, the CW signal of MF R-Mode is continuous signal, making it difficult to separate the direct signal and skywave \cite{Johnson2017:initial}.

In this study, we analyzed the positioning accuracy of MF DGNSS R-Mode in daytime and nighttime based on preliminary data from the MF R-Mode testbed in Korea under development. 
For the analysis, the data were collected in Daesan and Daejeon, Korea, and the received signals were transmitted from Chungju, Korea.

\section{MF R-Mode Skywave Model}

An MF DGNSS reference station transmits a signal by modulating the GNSS correction information using MSK in the 285--325 kHz range. 
When the amplitude of the signal is \textit{A}, the bit interval is \textit{T}, carrier frequency is $f_{c}$, phase offset is $\Phi$, and bit stream is $\pm1$, the MSK signal is expressed by (\ref{eqn:msk}) in \cite{Pasupathy1979:minimum}.
\begin{equation} 
\label{eqn:msk}
s_{\mathrm{msk}}(t)=A \mathrm{cos} \left( 2\pi f_{c}t \pm\frac{\pi t}{2T}+\Phi \right).
\end{equation}
One of the existing MF DGNSS R-Mode implementations adds two continuous waves at the frequencies of $f_{c} \pm250$ Hz to the MF DGNSS signal for ranging. 
With amplitude \textit{B} and phase offset $\Phi$, the continuous wave signals can be expressed as \cite{Grundhofer:phase}
\begin{equation} 
    \label{eqn:CW}
    \begin{split}
        &s_{\mathrm CW1} \left( t \right) = B_{\mathrm CW1} \cos \left( 2\pi \left( f_{c} - 250 \right) t +\Phi_{\mathrm CW1} \right)\\
        &s_{\mathrm CW2} \left ( t \right) = B_{\mathrm CW2} \cos \left( 2\pi \left( f_{c} + 250 \right) t +\Phi_{\mathrm CW2} \right).
    \end{split}
\end{equation}
The combined signal of (\ref{eqn:msk}) and (\ref{eqn:CW}) is transmitted as an MF R-Mode signal and is represented by the following equation:
\begin{equation} 
    \label{eqn:R_mode}
    \begin{split}
        s(t) =&s_\mathrm{msk}(t)+s_{CW1}(t)+s_{CW2}(t) \\
        =& A \cos \left ( 2\pi f_{c}t \pm\frac{\pi t}{2T}+\Phi \right ) \\
        +& B_{\mathrm CW1} \cos \left( 2\pi \left( f_{c} - 250 \right) t +\Phi_{\mathrm CW1} \right) \\
        +& B_{\mathrm CW2} \cos \left( 2\pi \left( f_{c} + 250 \right) t +\Phi_{\mathrm CW2} \right).
    \end{split}
\end{equation}

The receiver receives a groundwave, where the transmitted signal propagates along the ground propagation path, and a skywave that is reflected by the ionosphere. 
The skywave can be modeled as a time-delayed and amplitude-scaled version of the groundwave \cite{Johnson2014:feasibility3}. 
According to the results of \cite{Johnson2014:feasibility3}, the R-Mode positioning error's lower bound was 10 m during daytime in the North Sea area; however, it was 100 m at nighttime, which is a 10 times reduction. 
Skywaves begin at sunset and last until sunrise; thus, adversely affecting the nighttime performance. 
The skywave model with time delay $t_d$ and attenuation factor $\alpha$ is expressed as follows \cite{Johnson2014:feasibility3}:
\begin{equation} 
    \label{eqn:skywave}
    r\left ( t \right )=s\left ( t \right )+\alpha s\left ( t-t_{d} \right ).
\end{equation}

\section{Measurement Data Analysis}

\begin{table}
    \centering
    \caption{RMS distance error (m) of CW1 and CW2 in daytime and nighttime (Daejeon)}
    \begin{tabular}{|c|c|c|}
        \hline
            & \textbf{Daytime} & \textbf{Nighttime} \\ \hline
        CW1 & 320095     & 427741       \\ \hline
        CW2 & 320060     & 427741       \\ \hline
    \end{tabular}
    \label{tab:dj_rmse_1}
\end{table}

\begin{table}
    \centering
    \caption{RMS distance error (m) of CW1 and CW2 in daytime and nighttime (Daesan)}
    \begin{tabular}{|c|c|c|}
        \hline
            & \textbf{Daytime} & \textbf{Nighttime} \\ \hline
        CW1 & 229602     & 343526       \\ \hline
        CW2 & 229618     & 343399       \\ \hline
    \end{tabular}
    \label{tab:ds_rmse_1}
\end{table}

\begin{figure}
    \centering
    \includegraphics[width=0.9\linewidth]{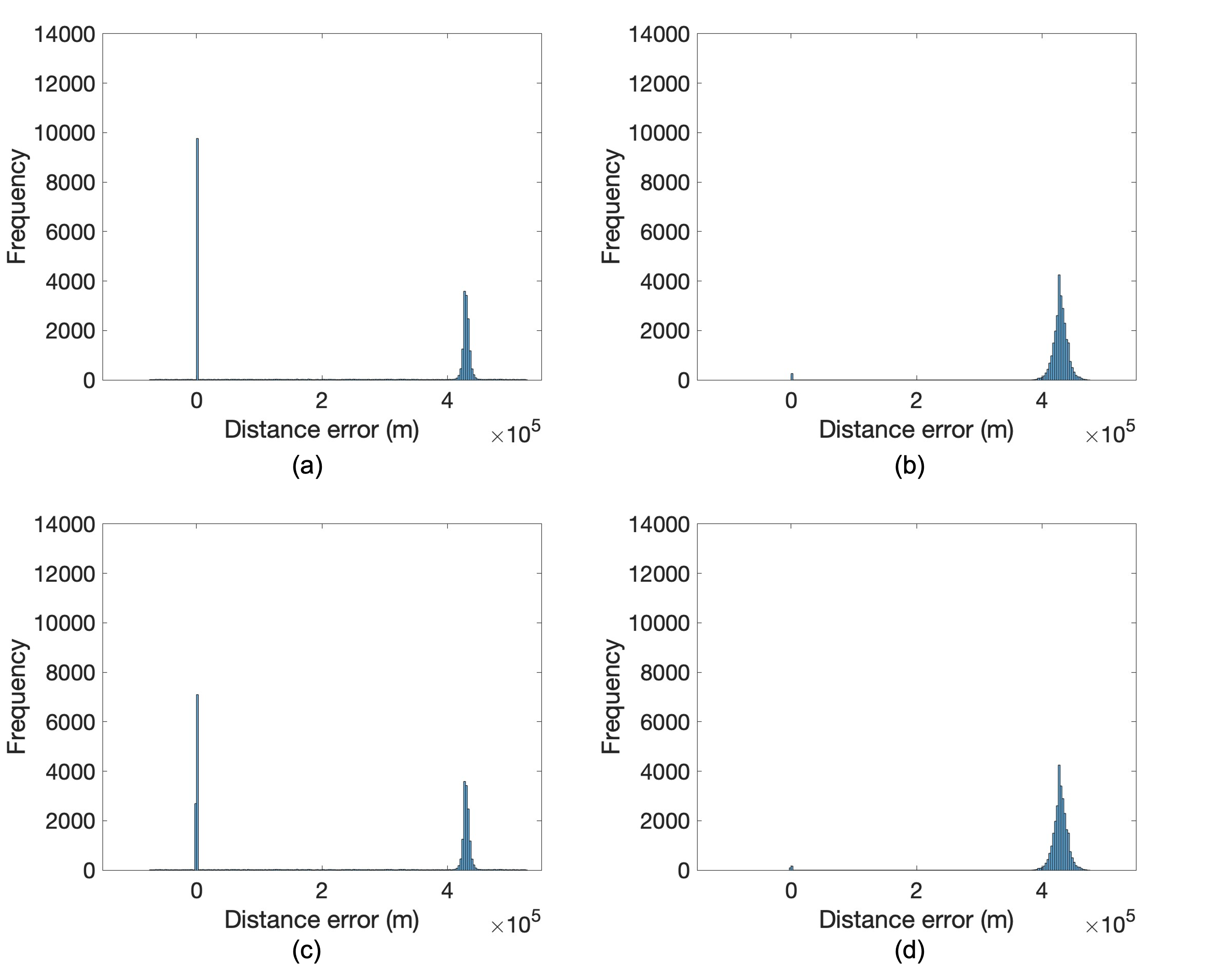}
    \caption{Distance error distribution during (a) daytime and (b) nighttime with CW1 signal, and (c) daytime and (d) nighttime with CW2 signal in Daejeon, Korea.}
    \label{fig:dj_rmse_1}
\end{figure}

\begin{figure}
    \centering
    \includegraphics[width=0.9\linewidth]{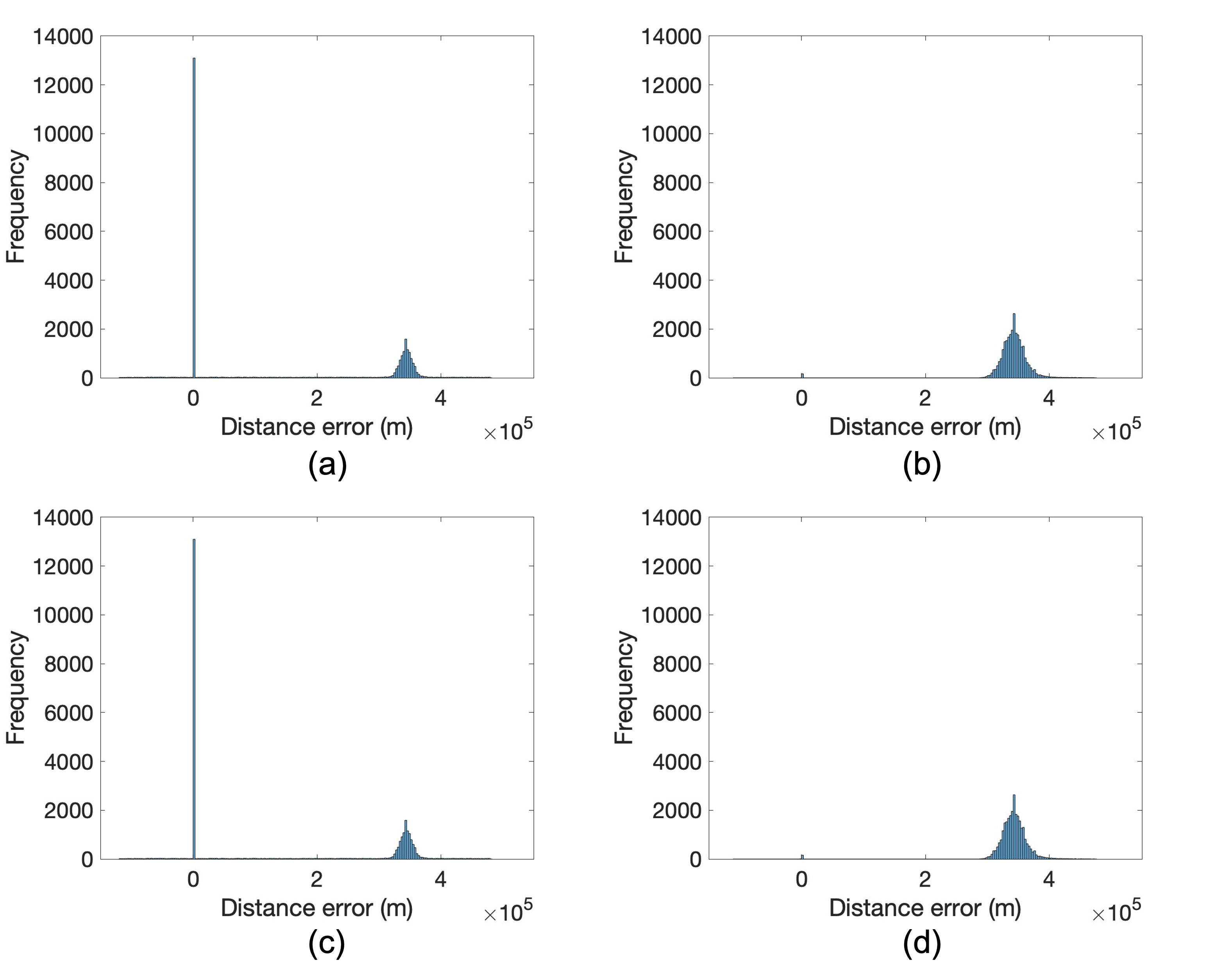}
    \caption{Distance error distribution during (a) daytime and (b) nighttime with CW1 signal, and (c) daytime and (d) nighttime with CW2 signal in Daesan, Korea.}
    \label{fig:ds_rmse_1}
\end{figure}

The R-Mode signal transmitter for our measurement campaign was in Chungju, Korea, and the receiver was in Daesan and Daejeon, Korea. 
The receiver used for this study is the Serco’s ``MFR-1a Medium Frequency R-Mode Receiver,'' which was also used in \cite{Johnson2020:R-Mode}. 
Signals were simultaneously collected from two receivers. 
The receiver provided the distance data used for the analysis. 
The transmission center frequency $f_c$ was 318 kHz, and CW1 and CW2 were 318.25 kHz and 317.75 kHz, respectively. 
Daytime refers to the time from 6 am to 6 pm, and the nighttime refers to the time from 6 pm to 6 am the next morning. 
The data was collected from 1:30 pm on April 21, 2022 to 11:30 am the next day. 
To investigate the effect of increased signal intensity on positioning accuracy, the intensity of CW signals was increased for three hours from 9 pm to 12 am.

The received data was analyzed in two parts. 
First, the distribution of distance measurement errors between daytime and nighttime was compared to determine the skywave effect at night. 
Second, data with increased CW signal intensity was compared with data without the increase from the nighttime measurements. 
The received signal with the signal-to-noise ratio (SNR) of 7 dB or greater was used for the analysis, which is the minimum standard for a maritime DGNSS beacon receiver prescribed by the International Electrotechnical Commission (IEC) \cite{IEC:International}.

\subsection{Daytime and Nighttime Comparison of MF DGNSS R-Mode Accuracy}
\label{sec:DN_comparision}

To equalize the amount of daytime and nighttime data, daytime data from 1:25 pm to 6 pm and 6 am to 9 am the next day, and nighttime data from 6 pm to 9 pm and 1:25 am to 6 am were used. 
The root mean square (RMS) errors of the distance measurements provided by the Serco receiver during daytime and nighttime in Daejeon and Daesan are listed in Tables \ref{tab:dj_rmse_1} and \ref{tab:ds_rmse_1}, respectively.
The histograms of the error distributions in Daejeon and Daesan are shown in Figs. \ref{fig:dj_rmse_1} and \ref{fig:ds_rmse_1}, respectively.

Regarding the distribution of distance errors in both Daesan and Daejeon data, it is evident that there is a high frequency of small distance errors during the daytime; however, at nighttime, the error is very large as it is rarely measured accurately. 
The distance error distributions around the very large errors in Figs. \ref{fig:dj_rmse_1} and \ref{fig:ds_rmse_1} are likely caused by wrong integer ambiguity resolution in the receiver.
Thus, the RMS distance error values in Tables \ref{tab:dj_rmse_1} and \ref{tab:ds_rmse_1} do not reflect the potential performance of the MF R-Mode with a correct integer value. 
Nevertheless, the different error distributions between daytime and nighttime are clearly shown in Figs. \ref{fig:dj_rmse_1} and \ref{fig:ds_rmse_1}.
Unlike \cite{Johnson2014:feasibility3}, where accuracy was measured at sea, in the testbed of our study, the propagation path was along the ground, which causes additional errors.

\subsection{MF DGNSS R-Mode Accuracy Analysis with Modified and Unmodified Signals}

From 9 pm to 12 am, the intensity of the CW signal was increased.
This signal with the increased intensity is called ``modified signal'' in this paper, and the original signal without the increased intensity is called ``unmodified signal.''
This three-hour data with the modified signal was compared with another three-hour data with the unmodified signal collected from 12 am to 3 am. 

\begin{figure}
    \centering
    \includegraphics[width=0.9\linewidth]{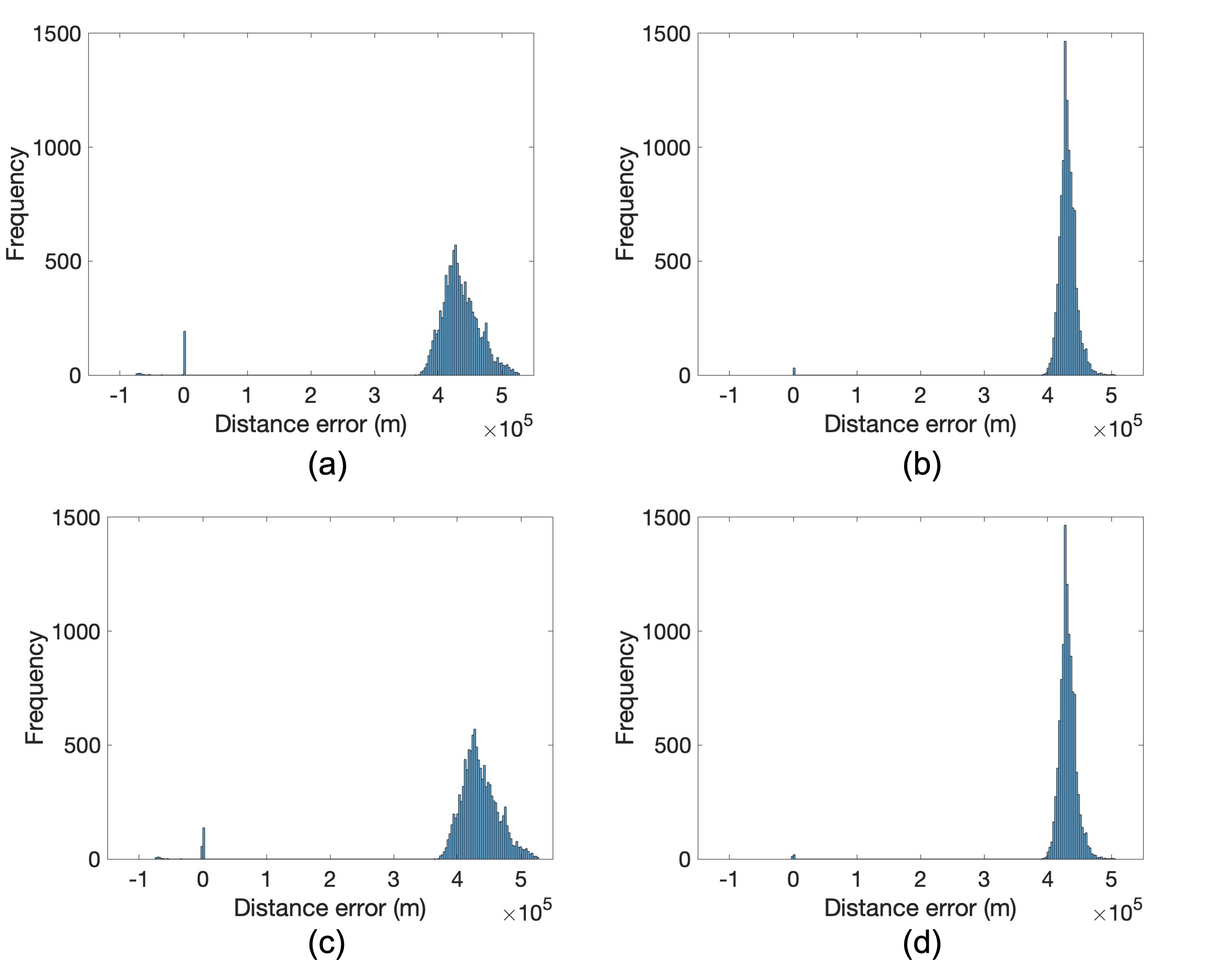}
    \caption{Range error distribution of (a) modified and (b) unmodified CW1 signal, and (c) modified and (d) unmodified CW2 signal in Daejeon, Korea.}
    \label{fig:dj_rmse_2}
\end{figure}

\begin{figure}
    \centering
    \includegraphics[width=0.9\linewidth]{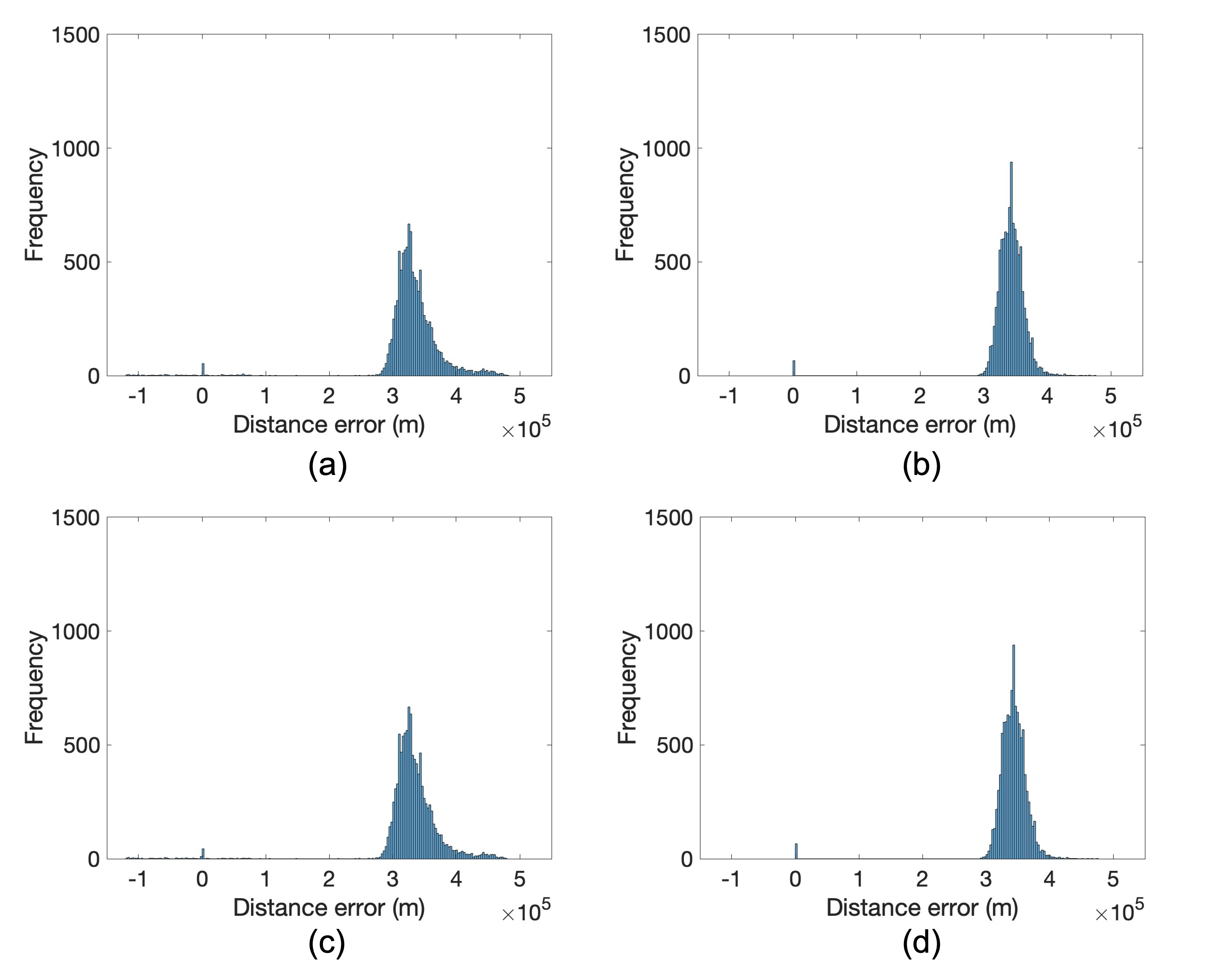}
    \caption{Range error distribution of (a) modified and (b) unmodified CW1 signal, and (c) modified and (d) unmodified CW2 signal in Daesan, Korea.}
    \label{fig:ds_rmse_2}
\end{figure}

Tables \ref{tab:SNR_dj} and \ref{tab:SNR_ds} lists the average SNR of the modified and unmodified signals according to CW1 and CW2.  
Similar to Section \ref{sec:DN_comparision}, we compared the distance error distributions measured in Daejeon and Daesan. 
The results are shown in Figs. \ref{fig:dj_rmse_2} and \ref{fig:ds_rmse_2}.
The distributions of the modified signal cases are closer to the zero, which implies more accurate distance measurements. 
Because a CW signal is used to measure distance, the stronger the received CW signal, the more accurate the measured distance. 
This suggests the possibility of increasing the nighttime accuracy by adjusting the CW signal strength.

\begin{table}
    \centering
    \caption{SNR (dB) of modified and unmodified CW1 and CW2 signals (Daejeon)}
    \begin{tabular}{|c|c|c|}
        \hline
            & \textbf{Modified} & \textbf{Unmodified} \\ \hline
        CW1 & 21.313     & 20.9757      \\ \hline
        CW2 & 21.877     & 21.252       \\ \hline
    \end{tabular}
    \label{tab:SNR_dj}
\end{table}

\begin{table}
    \centering
    \caption{SNR (dB) of modified and unmodified CW1 and CW2 signals (Daesan)}
    \begin{tabular}{|c|c|c|}
    \hline
        & \textbf{Modified} & \textbf{Unmodified} \\ \hline
    CW1 & 16.331     & 14.594       \\ \hline
    CW2 & 16.774     & 15.152       \\ \hline
    \end{tabular}
    \label{tab:SNR_ds}
\end{table}



\section{Conclusion}
\label{sec:Conclusion}

In this study, we analyzed the signals collected in Daesan and Daejeon to investigate the skywave effects on the MF R-Mode. 
The analysis was conducted by comparing the distance error distribution based on time of the day (daytime or nighttime) and the signal intensity of the CW signal. 
The results show that the accuracy of measuring the distance during daytime was higher than the case of  nighttime, which is due to skywave.
Furthermore, this study shows that higher CW signal strength can potentially reduce the influence of skywave.


\section*{Acknowledgment}

This research was conducted as a part of the project titled ``Development of integrated R-Mode navigation system [PMS4440]'' funded by the Ministry of Oceans and Fisheries, Republic of Korea (20200450).

\bibliographystyle{IEEEtran}
\bibliography{mybibfile, IUS_publications}

\vspace{12pt}

\end{document}